\documentstyle[pra,aps]{revtex}
          \begin{document}
          \draft
          \title{Quantum Robots and Quantum Computers}
          \author{Paul Benioff\\
           Physics Division, Argonne National Laboratory \\
           Argonne, IL 60439 \\
           e-mail: pbenioff@anl.gov}
           \date{\today}

          \maketitle
          \begin{abstract}  
Validation of a presumably universal theory, such as quantum
mechanics, requires a quantum mechanical description of systems 
that carry out theoretical calculations and sytems that carry out
experiments. The description of quantum computers is under active
development.  No description of systems to carry out experiments
has been given.  A small step in this direction is taken here by
giving a description of quantum robots as mobile systems with on
board quantum computers that interact with different
environments. Some properties of these systems are discussed.  A
specific model based on the literature descriptions of quantum
Turing machines is presented. 
 \end{abstract}

\section{Introduction}

Much of the impetus to study quantum computation, either as 
networks of quantum gates \cite{Vedraletal,Barencoetal} (See
\cite{EkJo} for a review) or as Quantum Turing Machines
\cite{Benioff8082,Benioff86,Deutsch85,Deutsch89,BenioffQBE}, is
based on the increased efficiency of quantum computers compared
to classical computers for solving some important problems
\cite{Shor,Grover}. Realization of this goal or use of quantum
computers to simulate other physical systems
\cite{Feynman82,Deutsch85} requires the eventual physical
construction of quantum computers.  However, as emphasized
repeatedly by Landauer \cite{Landauer}, there are serious
obstacles to such a physical realization. 

There is, however, another reason to study quantum computers that
is less dependent on whether or not such machines are ever built.
It is based on the fact that testing the validity of a physical
theory such as quantum mechanics requires the comparison of
numerical values calculated from theory with experimental
results.  If quantum mechanics is universally valid (and there is
no reason to assume otherwise), then both the systems that carry
out theoretical calculations and the systems that carry out
experiments must be described within quantum mechanics.  It
follows that the systems that test the validity of quantum
mechanics must be described by the same theory whose validity
they are testing.  That is quantum mechanics must describe its
own validation to the maximum extent possible \cite{PerZur}.

Because of these self referential aspects, limitations in
mathematical systems expressed by the G\"{o}del theorems lead one
to expect that there may be interesting questions of self
consistency and limitations in such a description.  Limitations
on self observation by quantum automata
\cite{Albert,Breuer,Peres} may also play a role here.

In order to investigate these questions it is necessary to have
well defined completely quantum mechanical descriptions of
systems that compute theoretical values and of systems that carry
out experiments.  So far there has been much work on quantum
computers.  These are systems that can, in principle at least,
carry out computation of theoretical values for comparison with
experiment.  However there has been no comparable development of
a quantum mechanical description of robots.  These are systems
that can, in principle at least, carry out experiments.

Another reason quantum robots are interesting is that it is
possible that they might provide a {\em very small} first step
towards a quantum mechanical description of systems that are
aware of their environment, make decisions, are intelligent, and
create theories such as quantum mechanics
\cite{Penrose,Stapp,Squires}. If quantum mechanics is universal,
then these systems must also be described in quantum mechanics to
the maximum extent possible.

The main point of this paper is that quantum robots and their
interactions with environments may provide a well defined
platform for investigation of many interesting questions
generated by the above considerations.  To this end some aspects
of quantum robots and their interactions with environments are
discussed in the next section. The close relation between quantum
robots and quantum computers is clear from the definition of a
quantum robot  as a mobile system consisting of an on board
quantum computer and needed ancillary systems that moves in and
interacts with an environment. 

A specific model of quantum robots plus environments is discussed
in Section \ref{SMQRE}.  The model, which is based on the
description of quantum Turing machines, describes the motion of a
quantum robot in an environment which is a 1-D lattice of qubits.
The overall model system consisting of a quantum robot plus
environment is considered to be isolated with dynamics given by a
time independent self adjoint Hamiltonian.  The model is in
essence also a slowed down version of a quantum Turing machine
which is described so that it can be easily reinterpreted as a
quantum robot interacting with an environment.  This
interpretation is facilitated by separation of the step operator,
defined in other work \cite{Deutsch85,BenioffQBE}, into two parts
describing action and computation phases.  

In the last section some similarities and differences between
quantum robots plus environments and quantum computers are
discussed. Quantum robots which function as quantum computers by
use of states of systems, that are a part of the environment, to
represent numbers are are seen to be limited in that there are
environments in which a satisfactory number representation is not
possible. Also the speculative possibility of a Church Turing
type hypothesis for the class of physical experiments is noted.

It must be emphasized that the language used in this paper to
describe quantum robots is carefully chosen to avoid any
suggestions that these systems are aware of their environment,
make decisions, carry out experiments or make measurements, or
have other properties characteristic of intelligent or conscious
systems.   The quantum robots described here have no awareness of
their environment and do not make decisions or measurements. 
Their description differs in detail only, from that used to
describe any other system in quantum mechanics.

It should be noted that some aspects of the ideas presented here
have already occurred in earlier work.  Physical operations have
been described as instructions for well-defined realizable and
reproducible procedures \cite{FouRan}, and quantum state
preparation and observation procedures have been described as
instruction booklets or programs for robots \cite{BenEks}. 
However these concepts were not described in detail and the
possibility of describing these procedures or operations quantum
mechanically was not mentioned. Also quantum computers had not
yet been described. More recently Helon and Milburn \cite{HeMi}
have described the use of the electronic states of ions in a
linear ion trap as an apparatus (and a quantum computer register)
to measure properties of vibrational states of the ions.  In
other work quantum mechanical Maxwell's demons have been
described \cite{Lloyd}. 

 Also there is much work on the interactions between quantum
computers and the environment.  However, these interactions are
considered as a source of noise or errors to be minimized or
corrected by use of quantum error correction codes \cite{Shor1}.
Here interactions between a quantum robot and the environment are
emphasized as an essential part of the overall system dynamics. 
Other work on environmental induced superselection rules
\cite{Zur,Joo} also emphasizes interactions between the
environment and a measurement apparatus that stabilize a selected
basis (the pointer basis) of states of the apparatus.

\section{Quantum Robots}
\label{QR}

Here quantum robots are considered to be mobile systems that have
a quantum computer on board and any other needed ancillary
systems.  Quantum robots move in and interact (locally) with
environments of quantum systems.  Since quantum robots are
mobile, they are limited to be quantum systems with finite
numbers of degrees of freedom.  

Environments consist of arbitrary numbers and type of systems
moving in 1-, 2-, or 3-dimensional spatial universes. The
component systems can have spin or other internal quantum numbers
and can interact with one another or be free.  Environments can
be open or closed.  If they are open then there may be systems
that remain for all time outside the domain of interaction with
the quantum robot that can interact with and establish
correlations with other environment systems in the domain on the
robot.  Quantum field theory may be useful to describe
environments containing an infinite number of degrees of freedom. 
To keep things simple, in this paper environments will be
considered to consist of systems in discrete space lattices
instead of in continuous space.

The quantum computer that is on board the quantum robot can be
described as a quantum Turing machine, a network of quantum
gates, or any other suitable model.  If it is a quantum Turing
machine, it consists of a finite state head moving on a finite
lattice of qubits.  The lattice can have distinct ends.  However
it seems preferable if the lattice is closed (i.e. cyclic).  If
the on board computer is a network of quantum gates then it
should be a cyclic network with many closed internal quantum wire
loops and a limited number of open input and output quantum wires
(narrow bandwidth).  Even though acyclic networks are sufficient
for the purposes of quantum computation \cite{Yao} cyclic ones
are preferable for quantum robots.  One reason is that
interactions between these networks and the environment are
simpler to describe and understand than those containing a large
number of input and output lines. Also the only known examples of
{\em very} complex systems that are aware of their environment
and are presumably intelligent, contain large numbers of internal
loops and internal memory storage. 

For the purposes of this paper the overall dynamics of a quantum
robot and its interactions with the environment is described in
terms of {\em tasks}.  A task for a quantum robot is equivalent
to a function associated with a quantum computer.  A quantum
robot carries out a task on some initial state of the environment
just as a quantum computer carries out a function computation on
a specified initial state.  

The goal of the task is to change the initial environmental state
into some final state with properties corresponding to the goal. 
An example of a task is "move each system in region $R$ 3 sites
to the right if and only if the destination site is unoccupied." 
Implementation of such a task requires specification of a path to
be taken by the quantum robot in executing the task. Some method
of determining when it is inside or outside of the specified
region and making appropriate movements must be available. In
this case if there are $n$ systems in region $R$ at locations
$x_{1},x_{2}, \cdots ,x_{n}$ in region $R$ then the initial state
of the regional environment, $ \vert \underline{x}\rangle =
\otimes_{j=1}^{n}\vert x_{j} \rangle$  becomes
$\otimes_{j=1}^{n}\vert x_{j}+3 \rangle$ provided all destination
sites are unoccupied.   

If the initial state of the regional environment is a linear
superposition of states $\psi
=\sum_{\underline{x}}c_{\underline{x}}\vert \underline{x}\rangle$
of n-system position states $\vert \underline{x}\rangle$ in $R$
then the final state of the regional environment is given in
general by a density operator even if all destination sites are
unoccupied.  This is a consequence of the fact that in general
the actions of the quantum robot introduce correlations between
the states of the robot systems and the different initial
environment component states $\vert \underline{x}\rangle$.  When
the task is completed on all components $\vert
\underline{x}\rangle$, the overall state of the robot plus
environment is given by a linear sum over robot regional
environment states of the form
$\sum_{\underline{x}}c_{\underline{x}}\theta_{\underline{x}}\vert
\underline{x}\rangle$.  Here $\theta_{\underline{x}}$ is the
final state of the quantum robot resulting from carrying out the
task on the regional environment in state
$\vert\underline{x}\rangle$.  Taking the trace over the robot
system variables gives the density operator form for the regional
environment state.

The above description shows that quantum robots can carry out the
same task on many different environments simultaneously.  This
can be done by use of an initial state of the quantum robot plus
environment that is a linear superposition of different
environment basis states.  For quantum computers the
corresponding property of carrying out many computations in
parallel has been known for some time \cite{Deutsch85}.  Whether
the  speedup provided by this parallel tasking ability can be
preserved for some tasks, as is the case for Shor's \cite{Shor}
or Grover's algorithms \cite{Grover} for quantum computers,
remains to be seen.

The above described task is an example of a reversible task. 
There are also many tasks that are irreversible. An example is
the task "clean up the region $R$ of the environment" where
"clean up" has some specific description such as "move all
systems in $R$ to some fixed pattern".  This task is irreversible
because many initial states of systems in $R$ are taken into the
same final state.  This task can be made reversible by storing
somewhere in the environment outside of $R$ a copy of each
component in some basis $B$ of the initial state of the systems
in $R$.  For example if $B=\{\vert \underline{x}\rangle\}$ and
$\sum_{\underline{x}}c_{\underline{x}}\vert \underline{x}\rangle$
is the initial state, then the copy operation is given by
$\sum_{\underline{x}}c_{\underline{x}}\vert \underline{x}\rangle
\vert \underline{0}\rangle_{cp} \longrightarrow
\sum_{\underline{x}}c_{\underline{x}}\vert \underline{x}\rangle
\vert \underline{x}\rangle_{cp}$.

This operation of copying relative to the states in some basis
avoids the limitations imposed by the no-cloning theorem
\cite{WooZur} because an unknown state is not being copied.  The
price paid is that copying relative to some basis introduces
branching into the process in that correlations are introduced
between the state of systems in the copy region and states of
systems in $R$. This is the quantum mechanical equivalent of the
classical case of making a calculation of a many-one function
reversible by copying and storing the input \cite{Bennett}.

In the above case carrying out the cleanup on the state
$\sum_{\underline{x}}c_{\underline{x}}\vert \underline{x}\rangle
\vert \underline{x}\rangle_{cp}$ corresponds to the operation
$\sum_{\underline{x}}c_{\underline{x}}\vert \underline{x}\rangle
\vert \underline{x}\rangle_{cp}\longrightarrow \vert
\underline{y}\rangle \sum_{\underline{x}}c_{\underline{x}}\vert
\underline{x}\rangle_{cp}$ where $\vert \underline{y}\rangle$ is
the clean up state for the region $R$. The overall process is
reversible as it can be described by the transformation
$\sum_{\underline{x}}c_{\underline{x}}\vert \underline{x}\rangle
\vert \underline{0}\rangle_{cp} \longrightarrow \vert
\underline{y}\rangle \sum_{\underline{x}}c_{\underline{x}}\vert
\underline{x}\rangle_{cp}$. If the final state of the quantum
robot depends on the initial state of the systems in region $R$,
then correlations remain and the overall transformation
corresponding to carrying out the cleanup task is given by 
$\sum_{\underline{x}}c_{\underline{x}}\vert \underline{x}\rangle
\vert \underline{0}\rangle_{cp}\theta_{i} \longrightarrow \vert
\underline{y}\rangle \sum_{\underline{x}}c_{\underline{x}}\vert
\underline{x}\rangle_{cp}\theta_{\underline{x}}$.  Here
$\theta_{i}$ and $\theta_{\underline{x}}$ are the initial and
final states of the quantum robot.

Each task is considered here to consist of a sequence of
computation and action phases.  The purpose of each computation
phase is to determine what action the quantum robot should take. 
The input to the computation carried out by the on board quantum
computer includes the local state of the environment and any
other pertinent information, such as the output of the previous
computation phase.  During a computation phase the robot does not
move or change the state of the environment.  It does change the
state of an on board ancillary system, the output system (o)
whose state determines the action taken following completion of
the computation.

During each action phase the quantum robot makes local changes in
the environment state or moves on the lattice. It can carry out
either or both of these types of steps.  Depending on the model
used, each action phase can consist of one step or several steps.
(The specific model described in the next section includes
multistep actions.)  Here one step consists of the robot moving
to at most an adjacent lattice site, or changing the state of the
environment in the neighborhood of the quantum robot, or both.
During an action phase the state of the (o) system, which
determines the action to be carried out,and the state of the on
board quantum computer, is not changed. Also the quantum robot
may or may not observe the local environment.  Examples of
actions that do not and do require local observations are "rotate
the qubit (as a spin « system) by an  angle $\phi$" and "rotate
the qubit by an angle $\phi$ only if it is in state $\vert
0\rangle$.  If the qubit is in state $\vert 1\rangle$ move to an
adjacent site."

The description of tasks carried out by quantum robots requires
the use of completion or halting flags to determine when
individual action and computation phases are completed as well as
when the overall task is completed.  Such flags are necessary if
the overall quantum robot plus environment dynamics is described
by a Hamiltonian because the unitarity of $e^{-iHt}$ requires
that system motion occurs somewhere even after the task is
completed.

Note that there are many examples of tasks that never halt. 
Nonhalting of tasks can arise from several sources.  The task may
consist of a nonterminating sequence of computation and action
phases. Or either a computation of an action phase may never
halt.  An example of an action that is multistep, does not halt,
and requires local environment interactions at each step is "move
along a string of $0s$ until a $1$ is found" carried out on a
lattice of $0s$ only in the direction of motion.

\section{A Specific Model of Quantum Robots plus Environments}
\label{SMQRE}
Here a specific model of quantum robots plus environments is
described that illustrates the above material. The close
relationship between quantum robots plus environments and quantum
computers is shown by the fact that the model also describes a
slowed down version of a quantum Turing machine.  In order to
have a model described entirely within quantum mechanics, the
overall system of quantum robot plus environment will be
considered to be isolated with dynamics given by a self adjoint
time independent Hamiltonian.  This avoids the presence of
external agents to turn on and off successive segments of a time
dependent Hamiltonian. Also, the model will be described using
information bearing degrees of freedom only.  The relevance of
this for the development of quantum computers has been noted by
Landauer{Land1}. 

The models are based on an expansion of quantum Turing machines
\cite{Benioff8082,Benioff86,Deutsch85,BeVa,BenioffQBE} to
describe models of quantum robots in a 1-D lattice qubit
environment.  The models also provide a natural decomposition of
each phase into one or more single steps.  The expansion is
straightforward as the models already describe a multistate head
moving on and interacting with a 1-D qubit lattice.
 
Models of quantum Turing machines consist of a 1-D finite  or
infinite lattice of qubits and a multistate head.  A computation
basis $B_{C}$ for the overall system of head and lattice consists
of the states $\{\vert l,j,\underline{s}\rangle\}$.   Here $\vert
l,j\rangle$ denotes the head internal state and lattice position
and $\vert \underline{s}\rangle=\otimes_{j=-\infty}^{\infty}\vert
\underline{s}_{j}\rangle$ denotes the state of the lattice qubit
systems.  For an infinite lattice $B_{C}$ is uncountably infinite
unless some tail condition is imposed on $\vert
\underline{s}\rangle$.  An example \cite{BenioffQBE} is that
$\vert \underline{s}_{j}\rangle \neq \vert 0\rangle$ for at most
a finite number of $j$. This tail condition applies to all states
in $B_{C}$.

Each QTM is described by a step operator $T$ acting on the
Hilbert space spanned by the basis $B_{C}$. $T$ is required to
satisfy locality and homogeneity conditions.  That is
\cite{Deutsch85},
 \begin{equation}
\langle l^{\prime},j^{\prime},\underline{s^{\prime}}\vert T\vert
l,j,\underline{s}\rangle = 0 \mbox{ if} \left[ \begin{array}{ll}
\underline{s^{\prime}},\underline{s} \mbox{ differ at positions
$\neq j$} \\ |j^{\prime}-j|>1 \end{array} \right. \label{Tloc}
\end{equation}
This expresses the locality condition in that single step changes
in the state of the lattice qubits are limited to the qubit at
the position of the head and the head can move at most one site
to the right or left.  Also the matrix element is independent of
the value of $j$ and depends on the difference $j^{\prime}-j$
only (homogeneity).

IF $T$ describes finite time interval steps as is done in some
models \cite{Deutsch85,BeVa} then $T$ is also required to be
unitary and iterations of $T$ or $T^{\dag}$ describe model
evolution.  This requirement is dropped in other work
\cite{BenioffQBE} in which $T$ is used to construct a Hamiltonian
according to Feynman's prescription \cite{Feynman},
\begin{equation}
H=K(2-T-T^{\dag}) \label{ham}
\end{equation}
where $K$ is a constant \cite{Benioff86,BenioffQBE}. As $H$ is
self adjoint and time independent, the finite time operator $e^{-iHt/\hbar}$ is unitary.

These models of quantum Turing machines can be changed into
models of quantum robots interacting with environments by
requiring the head $h_{1}$ to consist of an on board quantum
Turing machine and three other ancillary systems, a memory system
(m), an output system (o), and a control qubit (c). The on board
quantum Turing machine consists of another head $h_{2}$ moving on
a closed (e.g. circular) track of $N$ qubits.   Figure 1 shows
the complete system.  The qubit lattice ${\cal L}_{1}$ is the
environment of the quantum robot and ${\cal L}_{2}$ is the on
board $N$ qubit lattice.  The location of $h_{1}$ on ${\cal
L}_{1}$ is marked by an arrow.

In this model changes of the head internal state, which occur in
a single step in quantum Turing machines, Eq. \ref{Tloc}, become
multistep computations carried out by the on board quantum
computer in each computation phase.  A computation basis for the
on board quantum Turing machine has states of the form $\vert
p,k,\underline{t}\rangle$ that show $h_{2}$ in internal state
$\vert p\rangle$ at site $k$ on ${\cal L}_{2}$ and the ${\cal
L}_{2}$ qubits in state $\vert \underline{t}\rangle
=\otimes_{l=1}^{N}\vert \underline{t}_{l}\rangle$.  The three
added systems are used to regulate and determine the actions and
computations of the quantum robot.  The memory (m) and output (o)
systems are each described by an $L$ dimensional Hilbert space.
The control system (c) is a qubit. A reference basis set
$B_{omc}$ for the three systems has the form $\vert
l_{1}\rangle_{m} \vert l_{2}\rangle_{o} \vert i\rangle_{c}$ where
$l_{1},\; l_{2} =0,1,\cdots , L-1$ and $i=0,1$.

The function of the three added systems (o), (m), and (c) is
based on the separation of the step operator $T_{QR}$ into the
sum of two operators:
\begin{equation}
T_{QR}=T_{a}+T_{c}. \label{Tsum}
\end{equation}
where $T_{a}$ and $T_{c}$ describe respectively  single steps of
actions and computations of the quantum robot.  The dynamics of
the systems is given by Eq. \ref{ham} with $T_{QR}$ replacing
$T$.

The on board quantum Turing machine begins a $T_{c}$ computation
with (o), (m), and (c) in state $\vert l_{2}\rangle_{o} \vert
l_{1}\rangle_{m} \vert 0\rangle_{c}$ where $\vert l_{2}\rangle$
and $\vert l_{1}\rangle_{m}$ are the respective output from and
input to the previous computation This state and the reference
basis state $\vert s\rangle$ (with $s=0,1$) of the ${\cal L}_{1}$
qubit at the quantum robot location are the inputs to the
computation.

The goal of each computation phase is the computation of a new
state $\vert l_{3}\rangle_{o}$ of the output system and a shift
of the input state of (o) to the memory system (m).  The overall 
change of (o), (m), and (c) can be represented by $\vert
l_{2}\rangle_{o}\vert l_{1}\rangle_{m}\vert
0\rangle_{c}\longrightarrow \vert l_{3}\rangle_{o} \vert
l_{2}\rangle_{m} \vert 1\rangle_{c}$ which represents a change of
the states of (o), (m), and (c) systems in a reference basis
$B_{omc}$.   The last step of the computation is the conversion
of the control qubit state from $\vert 0\rangle$ to $\vert
1\rangle$ as $T_{c}$ is active only if the control qubit is in
state $\vert 0\rangle$.

This description applies to those computation phase operators
such that both $T_{c}$ and $T_{c}^{\dag}$  take states of
$B_{omc}$ into states of $B_{omc}$.  If either $T_{c}$ or
$T_{c}^{\dag}$ are such that iteration of these operators takes
states of $B_{omc}$ into linear superpositions of states in
$B_{omc}$, then branchings or entanglements are introduced.  

When the computation is finished the robot carries out the action
described by $T_{a}$.  The input to $T_{a}$ is the state $\vert
l_{3}\rangle_{o}\vert l_{2}\rangle_{m}\vert 1\rangle_{c}$. The
state $\vert l_{3}\rangle_{o}\vert l_{2}\rangle_{m}$  determines
which action the robot will carry out and the state $\vert
1\rangle_{c}$ activates the action phase of the robot.  If the
quantum robot completes the action, then the last step of $T_{a}$
is to change the control qubit state from $\vert 1\rangle$ back
to $\vert 0\rangle$. The (m) and (o) states are unchanged
throughout the action provided the states belong to a refernece
basis such as $B_{omc}$.  This restriction and that given above
for $T_{c}$ for the ${\cal L}_{1}$ qubit state avoid the
limitations of the no cloning theorem \cite{WooZur}.  This
completes the cycle as $T_{c}$ becomes active again.

The above description shows that the time evolution of the
quantum robot proceeds by alternating computing and action phases
each containing $\geq 1$ step or iteration of $T_{a}$ or $T_{c}$. 
One way to ensure that this proceeds smoothly is to require that,
except for the memory, output, and control systems, the terminal
state of the quantum Turing machine for one computation phase be
the same as the initial state for the next computation phase. For
example let $T_{c}$ begin and end a computation with  $\vert
p,k,\underline{t}\rangle =\vert 0,0,\underline{0}\rangle$.  The
main function of the ${\cal L}_{2}$ qubit lattice is as a scratch
pad for any calculation.  If part of it is set aside for added
memory or for input information, the above conditions on the
initial and final states would be changed to accomodate this.

The types and properties of possible actions that the quantum
robot can carry out in an action phase depend on the model being
considered. They can be either single step ($h_{1}$ motion at
most one ${\cal L}_{1}$ site) or multistep ($h_{1}$ motion of
several sites).  They also may or may not be mediated by
observations of the environment. For instance the action "rotate
the qubit by the angle $\phi$" is a single step action that
requires no observation. It applies to the qubit at the quantum
robot location whatever its state is. The multistep action "move
along a chain of $0s$, changing each qubit state $\vert 0\rangle$
to $a\vert 0\rangle+b\vert 1\rangle$ until a $1$ is encountered" 
requires observation of the environmental qubit at each
successive robot location to see if it is a $1$ or a $0$. 

The descriptions and requirements given above can be given in
terms of conditions that both $T_{a}$ and $T_{c}$ should satisfy.
$T_{c}$ is related to an operator $\tilde{T_{c}}$ defined on the
Hilbert space spanned by the basis set $\{\vert
p,k,t,s,l_{1},l_{2},i\rangle\}$. Here $\vert t\rangle$ and $\vert
s \rangle$ denote the states of the qubits at the locations of
$h_{2}$ on ${\cal L}_{2}$ and $h_{1}$ on ${\cal L}_{1}$
respectively, and $\vert i\rangle$ denotes the state of (c).  Let
the states $\vert \theta^{\prime}\rangle$ and $\vert \theta
\rangle$  denote respectively the states $\vert p^{\prime},
k^{\prime},s^{\prime},
l_{1}^{\prime},l_{2}^{\prime},i^{\prime}\rangle$ and $\vert p,k,
s,l_{1}, l_{2},i\rangle$.  One has
\begin{equation}
\langle \theta^{\prime},\underline{t^{\prime}}\vert T_{c}\vert
\theta,\underline{t}\rangle =  \langle \underline{t^{\prime}}
_{\neq k} \vert \underline{t}_{\neq k}\rangle \langle
\theta^{\prime},\underline{t^{\prime}}_{k} \vert
\tilde{T_{c}}\vert \theta, \underline{t}_{k}\rangle \label{T2def}
\end{equation}
Here $\langle \underline{t^{\prime}}_{\neq k}\vert
\underline{t}_{\neq k}\rangle$ denotes the product for all qubits
in ${\cal L}_{2}$ not at position $k$.  This condition states
that ${\cal L}_{2}$ qubit changes are limited to the qubit at the
location of $h_{2}$.

The operator  $\tilde{T}_{c}$ satisfies the conditions
\begin{equation}
\tilde{T_{c}}  =  \sum_{k,k^{\prime}}^{\prime}\sum_{s}
P_{k^{\prime}}P_{s}\tilde{T_{c}}P_{s}P_{k}P^{c}_{0}
\label{tT2def}
\end{equation}
where $P^{c}_{1}=\vert 1\rangle_{c}\langle 1\vert$ and
$P_{k}=\vert k\rangle \langle k\vert$.  The prime on the
$k,k^{\prime}$-sum means that it is limited to values for which
$|k^{\prime}-k|=0,1$. Also the values of the matrix elements of
$\tilde{T_{c}}$ depend on the difference $k^{\prime}-k$ and not
on the value of $k$.  

The equation states that when $T_{c}$ is active the state $\vert
s\rangle$ of  the qubit at the location of $h_{1}$ is not
changed.  This is expressed by the requirement that $T_{c}$ is
diagonal in the projection operator $P_{s}=\vert s\rangle \langle
s\vert $. Also single step motions of $h_{2}$ are limited to at
most one site on ${\cal L}_{2}$ and $\tilde{T_{c}}$ is active
(nonzero) only when the control qubit is in state $\vert
0\rangle$. 

The operator $T_{a}$ describes actions of the quantum robot
$h_{1}$ on ${\cal L}_{1}$. It is active in the Hilbert space
spanned by the basis $\{\vert
l_{2},l_{1},i,j,\underline{s},\rangle\}$ where $\vert
l_{2}\rangle,\; \vert l_{1}\rangle$, and $\vert i\rangle$ are the
respective states of the (o), (m), and (c) systems, and $j$ is
the position of $h_{1}$ on ${\cal L}_{1}$.  $T_{a}$ satisfies,
\begin{eqnarray}
\langle \phi^{\prime},j^{\prime}, \underline{s^{\prime}} \vert
T_{a}\vert \phi ,j,\underline{s}\rangle & = & \langle
\underline{s^{\prime}}_{\neq j}\vert \underline{s}_{\neq
j}\rangle \delta_{\underline{s^{\prime}}_{j},s^{\prime}}
\delta_{\underline{s}_{j},s} \nonumber \\
& & \mbox{} \times \langle \phi^{\prime},j^{\prime},s^{\prime}
\vert \tilde{T_{a}} \vert \phi ,j,s\rangle \label{T1def}
\end{eqnarray}
where $\vert phi \rangle = \vert l_{2},l_{1},i\rangle$. This
condition states that changes in ${\cal L}_{1}$ qubits are
limited to the qubit at the location of the quantum robot.

The operator $\tilde{T_{a}}$ satisfies the conditions
\begin{equation}
\tilde{T_{a}}  =  \sum_{j^{\prime},j}^{\prime}
\sum_{l,_{1},l_{2}}P_{j^\prime}P_{l_{1}}^{m}P_{l_{2}}^{o}\tilde{T_{a}}P_{l_{2}}^{o}P_{l_{1}}^{m}P_{j} P^{c}_{1} 
 \label{tT1def}
\end{equation}
The equation states that $\tilde{T_{a}}$ is diagonal in the
$B_{omc}$ basis for the (o) and (m) systems only and thus does
not change the (o) and (m) states provided they are in the
$B_{omc}$ basis.  As noted this avoids the limitations of the no
cloning theorem \cite{WooZur}. Also $T_{a}$ is active only when
$|j^{\prime}-j|\leq 1$ and the control system is in state $\vert
1\rangle$.

To avoid complications, the need for history recording has not
been discussed. Both the computation and action phases may need
to record some history. For example when $T_{c}$ is active, the
change $\vert l_{2}\rangle_{o} \vert l_{1}\rangle_{m}
\longrightarrow \vert l_{3}\rangle_{o} \vert l_{2}\rangle_{m}$
requires history recording if the change is not reversible. 
Where records are stored (on $h_{1}$ or in the environment)
depends on the model. Also the task carried out by the quantum
robot may not be reversible unless the initial environment is
copied or recovered.

Initial and final states for the starting and completion of 
tasks need to be described. For example at the outset the memory,
output, and control systems might be  in the state $\vert \vert
0\rangle_{m} \vert l_{i}\rangle_{o}\vert 0\rangle_{c}\vert
0\rangle_{c}$ and the environment would be in some suitable
initial state.   The process begins with the on board quantum
computer active.

Completion of a task could be described by designating one or
more states $\vert l_{f}\rangle$ as final output states and
arranging matters so that the action of $T_{a}$ based on any of
these states moves the quantum robot along ${\cal L}_{1}$ with no
changes in the environment state.  As is the case for
computation, continued motion of some type is necessary for any
reversible process.  

As noted the specific model of a quantum robot plus environment
is in essence a slowed down version of a quantum Turing machine. 
Each step of a quantum Turing machine is replaced by a multistep
quantum computation followed by a single action step.  This
replacement raises the question whether the model quantum robots
plus environments are as powerful when used as quantum computers
as the original quantum Turing machines.

This question is open.  However one can show that if the
dependence of $T_{a}$ on the memory states is removed, then the
specific models of quantum robots plus environments with dynamics
given by Eq. \ref{Tsum} are weaker as quantum computers than
quantum Turing machines described by Eq. \ref{Tloc}. To see this
let $T$ be such that $T\vert l,j,\underline{s}\rangle = \vert
l_{1},j_{1},\underline{s_{1}}\rangle$ and $T\vert
l^{\prime},j,\underline{s}\rangle= \vert l_{1}, j_{2},
\underline{s_{2}}\rangle$ where $l\neq l^{\prime},\; j_{1}\neq
j_{2}$, and $\underline{s_{1}} \neq \underline{s_{2}}$. 
Modelling this with a quantum robot requires that $T_{a}$ depend
on the memory state.  This is the reason that both the (o) and
(m) systems are present in the model.  However this dependence
can be excluded if desired as there is no a priori reason why
quantum robots plus environments, when functioning as quantum
computers, should be as powerful as quantum Turing machines.

\section{Discussion}
\label{Disc}
The model described in the last section raises the question "Are
there any real differences between quantum robots interacting
with environments and quantum computers?" Here it it will be seen
that the answer is that there are real differences. To begin with
one notes that quantum robots plus environments can function as
quantum computers in two ways. One obvious way is by use of the
quantum computer on board the robot as a stand alone quantum
computer with no environment-robot interactions needed.  The
other way uses states of a collection $C$ of systems in the
environment to construct a k-ary representation of numbers with
$k\geq 2$. In this case the quantum robot and the systems in $C$
are the quantum computer with interactions between the quantum
robot and the systems in $C$ generating the steps in the quantum
computation.  The fact that the quantum robot includes an on
board quantum computer is not relevant provided one is only
interested in using the system  (quantum robot plus $C$) as a
quantum computer. In the models described in section \ref{SMQRE}
the fact that states of the environment as a lattice of qubits
can be used to represent numbers is incidental to consideration
of the system as a quantum robot plus environment.\footnote{It
may be possible to carry these ideas over to quantum computers
represented by networks of quantum gates.  In this case the
network may be reinterpreted as a community of motionless, very
simple quantum robots interacting with an environment of moving
qubits.  Here also the collection $C$ is the whole environment. 
Such an interpretation depends on considering a single quantum
gate as a very simple quantum computer.}

This equivalence between quantum robots plus environments and
quantum computers holds only for those environments containing
collections $C$ of systems as described above.  However there are
many environments that do not contain any such collection $C$ of
systems.  Examples include environments of interacting moving
quantum systems or environments of systems whose only posssible
number representation is unary.  An example of the latter is a
single spinless system on a 1-D space lattice.  Here the number
representation is given by the distance or number of sites from
an origin to the system on the lattice.  The head interacting
with this environment does not correspond to an efficient quantum
computer.  On the other hand this example, and the example of
moving interacting systems are acceptable environments for a
quantum robot.  Note that no requirement of efficiency is imposed
for quantum robots.

The above illustrates one of the differences between quantum
robots interacting with environments and quantum computers.
Another difference is a matter of emphasis.  For quantum robots
the emphasis is on the quantum robot interacting with and
changing the state of external systems that are not part of the
robot. Because the systems are external to the robot and are part
of the environment, problems with self observation by quantum
automata \cite{Albert,Breuer,Peres} do not arise.

For quantum computers the interactions between the systems whose
states are used to represent numbers and other computer
components are internal to and an essential part of the computer. 
Effects of external systems are to be minimized or corrected for
\cite{Shor1}.  This is the case whether these systems are part of
the quantum robot (as the head) or are part of the environment. 
Here problems with self observation may arise because systems
that are external to the system as a quantum robot plus
environment are now internal to and part of the quantum computer.

In conclusion the following speculative ideas may be worth
considering.  The close connection between quantum computers and
quantum robots interacting with environments suggests that the
class of all possible physical experiments may be amenable to
characterization just as is done for the computable functions by
the Church-Turing hypothesis \cite{Turing,Deutsch85,Nielsen}. 
That is there may be a similar hypothesis for the class of
physical experiments.

The description of tasks carried out by quantum robots (Section
\ref{QR}) lends support to this idea in that there may be an
equivalent Church Turing hypothesis for the collection of all
tasks that can be carried out.  The earlier work that
characterizes physical proceedures as collections of instructions
\cite{FouRan,Eks}, or state preparation and observation
proceedures as instruction booklets or programs for robots
\cite{BenEks} also supports this idea.  On the other hand much
work needs to be done to give a precise characterization of
physical experiments, if such is indeed possible.

\section*{Acknowledgements}
This work is supported by the U.S. Department of Energy, Nuclear 
Physics Division, under contract W-31-109-ENG-38.

\begin{center}
FIGURE CAPTION
\end{center}

Figure 1.  A Model of a Quantum Robot and its Environment. The
environment is an infinite 1-D lattice ${\cal L}_{1}$ of qubits. 
The quantum robot $h_{1}$ consists of an on board Quantum Turing
machine, finite state memory (m) and output (o) systems, and a
control qubit (c). The on board QTM consists of a finite closed
lattice ${\cal L}_{2}$ of qubits and a finite state head $h_{2}$
that moves on ${\cal L}{_2}$. The position of the quantum robot
$h_{1}$ on the environment lattice ${\cal L}_{1}$ is shown by an
arrow.


\begin{thebibliography}{99}

\bibitem{Vedraletal}
V. Vedral, A. Barenco, and A. Ekert, Phys. Rev. A {\bf 54} 147
(1996)

\bibitem{Barencoetal}
A. Barenco, C. H. Bennett, R. Cleve, D. P. DiVincenzo, N.
Margolus, P. Shor, T. Sleator, J. A. Smolin, and H. Weinfurter,
Phys. Rev. A {\bf 52} 3457 (1995)

\bibitem{EkJo}
A. Ekert and R. Jozsa, Revs. Modern Phys. {\bf 68} 733 (1996)

\bibitem{Benioff8082}
P. Benioff, Jour. Stat. Phys. {\bf 22} 563 (1980); Phys. Rev.
Letters {\bf 48} 1581 (1982).

\bibitem{Benioff86}
P. Benioff, Ann. NY Acad. Sci. {\bf 480} 475 (1986)

\bibitem{Deutsch85}
D. Deutsch, Proc. Roy. Soc. (London) A {\bf 400} 997 (1985).

\bibitem{Deutsch89}
D. Deutsch, Proc. Roy. Soc. (London) A {\bf 425} 73 (1989)

\bibitem{BenioffQBE}
P. Benioff, Phys. Rev. A {\bf 54} 1106 (1996); Phys. Rev. Letters
{\bf 78} 590 (1997)

\bibitem{Shor}
P. Shor, in {\it Proceedings of the 35th Annual Symposium on the
Foundations of Computer Science}, edited by S. Goldwasser (IEEE
Computer Society, Los Alamitos, CA 1994), p. 124; Siam Jour.
Comput. {\bf 26}, 1481 (1997).

\bibitem{Grover}
L.K.Grover, in {\it Proceedings of 28th Annual ACM Symposium on
Theory of Computing} ACM Press New York 1996, p. 212; Phys. Rev.
Letters, {\bf 78} 325 (1997); G. Brassard, Science {\bf 275} 627
(1997).

\bibitem{Feynman82}
R. P. Feynman, International Jour. of Theoret. Phys. {\bf 21} 467
(1982)

\bibitem{Landauer}
R. Landauer, Physics Letters A {\bf 217} 188 (1996); Phil. Trans.
R. Soc. Lond. A {\bf 353} 367 (1995); Physics Today {bf 44} 23
(1991) May; IEEE Transactions on Electron Devices, {\bf 43} 1637
(1996).

\bibitem{PerZur}
A. Peres and W. Zurek, Amer. Jour. Phys. {\bf 50} 807 (1982).

\bibitem{Albert}
D. Albert, Physics Letters {\bf 98A} 249 (1983); Philosophy of
science {\bf 54} 577 (1987); {\it The Quantum Mechanics of Self-measurement} in {\bf Complexity, Entropy and the Physics of
Information}, proceedings of the 1988 workshop in Santa Fe, New
Mexico, 1989, W. Zurek, Ed.  Addison Wesely Publishing Co. 1990.

\bibitem{Breuer}
T. Breuer, Philos. Science {\bf 62} 197 (1995).

\bibitem{Peres}
Phys. Letters, {\bf A101} 249 (1984).

\bibitem{Penrose} 
R. Penrose, {\it The Emperor's New Mind}, Penguin Books, New
York, 1991.

\bibitem{Stapp}
H. P. Stapp, {\it Mind, Matter, and Quantum Mechanics}, Springer
Verlag, Berlin 1993.

\bibitem{Squires}
E. Squires, {\it Conscious Mind in the Physical World} IOP
Publishing, Bristol England, 1990

\bibitem{FouRan}
C. H. Randall and D. J. Foulis, Amer. Math. Monthly, {\bf 77} 363
(1970); D. J. Foulis and C. H. Randall, Jour. Math. Phys., {\bf
13} 1667 (1972).

\bibitem{BenEks}
P. Benioff and H. Ekstein, Phys. Rev. D {\bf 15} 3563, (1977);
Nuovo Cim. {\bf 40 B} 9 (1977).

\bibitem{HeMi}
C. D. Helon and G. J. Milburn, {\it Quantum Measurements with a
 Quantum Computer}, Los Alamos Archives preprint, quant-ph/9705014.

\bibitem{Lloyd}
S. Lloyd, Phys. Rev. A {\bf 56} 3374 (1997).

\bibitem{Shor1}
P. W. Shor, Phys. Rev. A {\bf 52} R2493 (1995); R. LaFlamme, C.
Miquel, J.P. Paz, and W. H. Zurek, Phys. Rev. Letters {\bf 77}
198 (1996); E. Knill and R. Laflamme, Phys. Rev. A {\bf 55}, 900
(1997); D. P. DiVincenzo and P. W. Shor, Phys. Rev. Letters {\bf
77} 3260 (1996).

\bibitem{Zur}
W. H. Zurek, Phys. Rev. D{\bf 24}, 1516 (1981); {\bf 26}, 1862
(1982); J. R. Anglin and W. H. Zurek, Phys. Rev. D {\bf 53}, 7327
(1996).

\bibitem{Joo}
E. Joos and H. D. Zeh, Z. Phys. B {\bf 59}, 223 (1985).

\bibitem{Yao}
A. Yao, in {\it Proceedings of the 34th Annual Symposium on
Foundations of Computer Science} (IEEE Computer Society, Los
Alamitos, CA, 1993), pp. 352-361.

\bibitem{Bennett}
C. H. Bennett, IBM Jour. Res. Dev. {bf 17}, 525 (1973).


\bibitem{Land1}
R. Landauer, {\it Zig-Zag Path to Understanding} in Proceedings
of the Workshop on Physics and Computation, PhysComp  94, Los
Alamitos: IEEE Computer Society Press, 1994.

\bibitem{BeVa}
E. Bernstein and U. Vazirani, in {\it Proceedings of the 1993 ACM
Symposium on Theory of Computing} (ACM, New York, 1993), pp 1-20.

\bibitem{Feynman}
R. P. Feynman, Optics News {\bf 11} 11 (1985); reprinted in
Foundations of Physics {\bf 16} 507 (1986).

\bibitem{WooZur}
W. K. Wootters and W. H. Zurek, Nature {\bf 299}, 802 (1982); H.
P. Yuen, Physics Letters {\bf 113A}, 405 (1986); H. Barnum, C. M.
Caves, C. A. Fuchs, R. Jozsa, and B. Schumacher, Phys. Rev.
Letters {\bf 76} 2818 (1996); L. M. Duan and G. C. Duo, Los
Alamos Archives, preprint no. quant-ph/9705018.


\bibitem{Turing}
A. Church, Am. Jour. Math. {\bf 58},345 (1936); A. M. Turing,
Proc. Lond. Math. Soc. 2 {\bf 42}, 230 (1936).

\bibitem{Nielsen}
M. A. Nielsen, Phys. Rev. Letters, {\bf 79} 2915 (1997); K
Svozil, {\it The Church-Turing thesis as a Guiding Principle for
Physics} Los Alamos Archives preprint quant-ph/9710052.

\bibitem{Eks}
H. Ekstein, Phys. Rev. {\bf 153}, 1397 (1967); {\bf 184}, 1315
(1969).

\end{thebibliography}
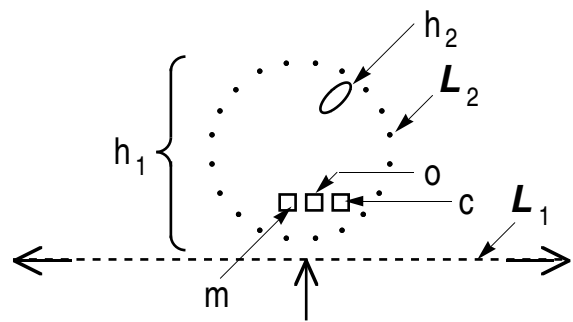
\end{document}